\documentclass[twocolumn,dvipsnames]{aastex62}
\usepackage{amsmath,amstext}
\usepackage{apjfonts}
\usepackage[figure,figure*]{hypcap}
\usepackage{ulem}
\usepackage{xspace}
\usepackage{lineno}
\usepackage{xfrac}
\usepackage{graphicx}
\usepackage{natbib}
\usepackage{multimedia}
\usepackage{amssymb}
\usepackage{amsmath}
\usepackage{array,booktabs}
\usepackage{mathptmx}
\usepackage{booktabs}
\usepackage{hyperref}
\usepackage{float}
\usepackage{todonotes}
\usepackage{multirow}

\newcommand{\erg}{\,{{\rm erg}}}

\newcommand{\msun}{\,{M_{\odot}}}
\newcommand{\Mtot}{\,{M_{\rm tot}}}
\newcommand{\Mthresh}{\,{M_{\rm thresh}}}
\newcommand{\mej}{\,{M_{\rm ej}}}
\newcommand{\RNS}{\,{{R_{\rm NS}}}}
\newcommand{\PNS}{\,{{P_{\rm NS}}}}
\newcommand{\PBH}{\,{{P_{\rm BH}}}}
\newcommand{\Eisog}{\,{E_{{\rm iso},\gamma}}}

\newcommand{\thmns}{\,{{t_{\rm HMNS}}}}
\newcommand{\s}{\,{{\rm s}}}

\newcommand{\ms}{\,{{\rm ms}}}

\newcommand{\km}{\,{{\rm km}}}

\newcommand{\Ye}{\,{Y_{\rm e}}}
\newcommand{\phim}{\,{\phi_{\rm M}}}

\shorttitle{A Unified Kilonova--GRB Model Establishes Neutron Stars as Short GRB Central Engines}
\shortauthors{Gottlieb, Metzger, Foucart \& Ramirez-Ruiz}

\begin{document}

\title{A Unified Model of Kilonovae and GRBs in Binary Mergers Establishes\\Neutron Stars as the Central Engines of Short GRBs}

    \author[0000-0003-3115-2456]{Ore Gottlieb}
	\email{ogottlieb@flatironinstitute.org}
	\affiliation{Center for Computational Astrophysics, Flatiron Institute, New York, NY 10010, USA}
    \affiliation{Department of Physics and Columbia Astrophysics Laboratory, Columbia University, Pupin Hall, New York, NY 10027, USA}

    \author[0000-0002-4670-7509]{Brian D. Metzger}
    \affiliation{Department of Physics and Columbia Astrophysics Laboratory, Columbia University, Pupin Hall, New York, NY 10027, USA}
    \affiliation{Center for Computational Astrophysics, Flatiron Institute, New York, NY 10010, USA}

    \author[0000-0003-4617-4738]{Francois Foucart}
	\affiliation{Department of Physics and Astronomy, University of New Hampshire, 9 Library Way, Durham, NH 03824, USA}

    \author[0000-0003-2558-3102]{Enrico Ramirez-Ruiz}
	\affiliation{Department of Astronomy and Astrophysics, University of California, Santa Cruz, CA 95064, USA}

\begin{abstract}

We expand the theoretical framework by \citet{Gottlieb2023e}, which connects binary merger populations with long and short binary gamma-ray bursts (lbGRBs and sbGRBs, respectively), incorporating kilonovae as a key diagnostic tool. We show that lbGRBs, powered by massive accretion disks around black holes (BHs), should be accompanied by bright, red kilonovae. In contrast, sbGRBs -- if also powered by BHs -- would produce fainter, red kilonovae, potentially biasing against their detection. However, magnetized hypermassive neutron star (HMNS) remnants that precede BH formation can produce jets with power ($P_{\rm NS} \approx 10^{51}\,{\rm erg\,s^{-1}}$) and Lorentz factor ($\Gamma>10$) likely compatible with sbGRB observations, and would result in distinctly bluer kilonovae, offering a pathway to identifying the sbGRB central engine. Recent modeling by \citet{Rastinejad2024} found luminous red kilonovae consistently accompany lbGRBs, supporting their origin in BH-massive disk systems, likely following a short-lived  HMNS phase. The preferential association of sbGRBs with comparably luminous kilonovae argues against the BH engine hypothesis for sbGRBs, while the bluer hue of these KNe provides additional support for an HMNS-driven mechanism. Within this framework, BH--NS mergers likely contribute exclusively to the lbGRB population with red kilonovae. Our findings suggest that GW170817 may, in fact, have been an lbGRB to on-axis observers. Finally, we discuss major challenges faced by alternative lbGRB progenitor models, such as white dwarf--NS or white dwarf--BH mergers and accretion-induced collapse forming magnetars, which fail to align with observed GRB timescales, energies, and kilonova properties.

\end{abstract}

\section{Introduction}

Gamma-ray bursts (GRBs) are often classified according to their progenitors: massive stars that collapse to Kerr black holes (BHs) are leading contenders for the engines of long-duration GRBs \citep{Woosley1993}, while mergers of double neutron star (NS) or BH--NS binaries result in short-duration GRBs \citep{Paczynski86,Eichler1989}. However, the discovery of radioactive decay-powered kilonova (KN) signals \citep{Metzger2010a,Roberts2011,Metzger2019} following two $ \sim 10\,\s $-long GRBs -- GRB 211211A \citep{Rastinejad2022, Troja2022, Yang2022, Zhang2022} and GRB 230307A \citep{Levan2023b, Sun2023, Yang2023} -- suggests a merger origin for these long-duration events, reshaping our understanding of GRB classification. These findings imply that some long bursts also arise from compact binary mergers, motivating the introduction of physically motivated sub-classes of long and short compact ``binary'' GRBs (dubbed lbGRBs and sbGRBs, respectively). They also challenge models predicting that jet launching in mergers occurs on much shorter timescales over which most of the merger debris accretes, typically $ t \ll 1\,\s $ \citep[e.g.,][]{Narayan+92,Lee2007}, rendering these $ 10\,\s $-long GRBs unexpectedly long.

\citet{Gottlieb2023e} established a theoretical framework linking binary merger populations to observed GRB populations as a result of variations in the mass of the post-merger accretion disk. They proposed that the duration of a BH-powered GRB jet is set by the time it takes the disk to reach a magnetically-arrested state \citep[MAD;][]{Narayan2003,Tchekhovskoy2011}, rather than the shorter time over which most of the {\it mass} accretes. After the disk becomes MAD, the jet power begins to decay as $ \sim t^{-2}$, and the transition to this phase can power the temporally ``extended'' prompt emission lasting up to minutes, observed in a high-fraction of GRBs (e.g., \citealt{Norris2006,Perley2009}). Building on this framework, \citet{Gottlieb2023e} showed that massive disks ($M_d\gtrsim0.1\,M_\odot$) around BHs, which form (for example) for large binary mass ratio $q\gtrsim1.2$ in NS--NS or $q\lesssim3$ in BH--NS mergers with rapidly rotating BHs \citep[see e.g.,][for a review]{Shibata2019}, will \emph{inevitably} produce such lbGRBs via the ``Blandford-Znajek'' mechanism \citep[hereafter BZ;][]{Blandford1977}.

Within this framework, the origin of sbGRBs remains less clear. {\it A priori}, two central engines remain contenders for powering the shortest bursts:
\begin{enumerate}
    \item BZ jets from BHs with low-mass disks, following a continuum from the massive-disk/lbGRB case; or
    \item Jets powered by a long-lived ($ 0.1\,\s \lesssim \thmns \lesssim 1\,\s $) magnetized NS remnant that precedes BH formation, a so-called ``hypermassive NS'' (HMNS) \citep[e.g.,][see however \citealt{Baumgarte2000} for the definition of an HMNS, which might differ from the properties of the NS remnant]{Murguia-Berthier2014,Ruiz2016,Ruiz2019,Ruiz2020,Ruiz2020b,Ciolfi2017,Ciolfi2019,Ciolfi2020,Mosta2020,Combi2023,Most2023,Most2023b,Bamber2024,deHaas2024,Kiuchi2024,Musolino2024}.
\end{enumerate}
While both scenarios could, in principle, account for the jetted activity of sbGRBs, they make distinct predictions for the merger's other electromagnetic counterpart: the KN \citep[e.g.,][]{Dessart2009,Metzger2014,Lippuner2017,Fahlman2018,Curtis2023,Curtis2024,Bernuzzi2024}, making it key to identifying the central engines of sbGRBs.

In this paper, we build on the unified model of \citet{Gottlieb2023e}, extending it to connect the diverse properties of merger GRBs not only to their progenitor binary populations but also to their KN counterparts. The paper is organized as follows: In \S\ref{sec:diskgrb}, we review the connection between disk mass and GRB characteristics. Building on this relationship, in \S\ref{sec:lbgrb}, we explain why lbGRBs originate from BHs with massive accretion disks. In \S\ref{sec:sbgrb}, we show that both BH-powered and HMNS-powered jets are consistent with sbGRB observations. In \S\ref{sec:kn}, we establish the connection between GRBs and KNe, and show that recent modeling of KNe associated with sbGRBs and lbGRBs \citep{Rastinejad2024} supports our prediction that lbGRBs are powered by BHs with massive accretion disks, and suggests that sbGRBs are likely powered by long-lived HMNSs. In \S\ref{sec:summary}, we conclude and discuss why other scenarios such as long-lived magnetars, white dwarf (WD)--NS, WD--BH mergers, and accretion-induced collapse (AIC) are challenged by both GRB and KN observations.

\section{Unified model for short and long compact binary GRB\MakeLowercase{s}}

We review the findings of \citet{Gottlieb2023e} for how the disk mass and magnetic flux on the BH jointly determine the jet power and duration. We divide the jet launching process into two phases: the prompt and extended emission. After summarizing how BH-powered jets naturally account for lbGRBs, we explore the implications for sbGRBs by generalizing the discussion to include NS-powered jets. 

\subsection{Black hole disk--GRB connection}\label{sec:diskgrb}

The time evolution of the BH mass accretion rate is governed by the total disk mass and its viscous spreading \citep[e.g.,][]{Lee2007}. The disk radius grows as $ r_d(t) \sim t^{2/3} $ while its mass decays as $ M_{\rm d}(t) \sim t^{-1/3} $ \citep{Metzger2008}. Consequently, the BH accretion rate evolves as \citep{Gottlieb2023e}
\begin{equation}\label{eq:Mdot}
    \dot{M}(t) \sim \frac{r_0}{r_d(t)}\frac{M_d(t)}{t} = \dot{M}(1\,\s)\left(\frac{t}{1\,\s}\right)^{-2}\,,
\end{equation}
where $ r_0 $ is the disk radius at formation, and the factor $ r_0/r_d $ accounts for suppression of the accretion rate reaching the BH due to outflows or convection in the disk \citep[e.g.,][]{Blandford1999}. Hereafter we assume the mass accretion rate decays as $ \dot{M}(t)\sim t^{-\alpha} $, where $ 1.5 \lesssim \alpha \lesssim 2 $ has been universally observed in numerical relativity and GRMHD simulations \citep[e.g.,][]{Fernandez2015,Fernandez2017,Fernandez2019,Christie2019,Metzger2021,Hayashi2022,Hayashi2023,Hayashi2024,Gottlieb2023e,Gottlieb2023b,Kiuchi2023b}.

The compact size and large vertical thickness of the post-merger accretion disk lead to the rapid advection of any magnetic flux present in the disk onto the BH. Therefore, the flux on the BH, $\Phi$, quickly saturates, resulting in a roughly constant jet power $ \PBH \sim \Phi^{2} \sim {\rm const.} $ during the prompt emission phase \citep[see also][]{Tchekhovskoy2015b}. By contrast, the dimensionless magnetic flux grows as
\begin{equation}\label{eq:phi}
    \phi(t) = \frac{\Phi(t)}{\sqrt{\dot{M}(t)r_g^2c}} \sim t^{\alpha/2} \sim t^{1}\,,
\end{equation}
where $ r_g $ is the BH gravitational radius. Once $\phi $ reaches $ \phim \approx 50 $ at $ t = t_{\rm MAD}$, magnetic stresses begin to dominate over accretion and the disk enters the MAD state \citep{Tchekhovskoy2011}. After the MAD transition, the BH cannot accept more magnetic flux; the dimensionless magnetic flux saturates at $\phi \approx \phim $, indicating that $\Phi$ begins to decay due to magnetic reconnection.

The power of a BH-driven jet, consistent with simulation results, is thus given by \citep{Gottlieb2023e}
\begin{equation}\label{eq:PBH}
	\PBH(t) = \eta_a\left[\frac{\phi(t)}{\phim}\right]^2\dot{M}(t)c^2 \sim 
	\begin{cases}
		{\rm const.} &\quad t<t_{\rm MAD}\, \\
		t^{-\alpha} &\quad t\geq t_{\rm MAD}\,,
	\end{cases} 
\end{equation}
where $ \eta_a \approx a^4+0.4a^2 $ \citep{Lowell2023} represents the BH spin-dependent efficiency, which remains constant in the post-merger phase since the BH mass $ M \gg M_d $, preventing any significant change in the BH spin $a$ as a result of disk accretion or jet launching. The first phase, characterized by constant jet power, corresponds to the main GRB prompt emission. A roughly constant jet power is consistent with observations that show weak temporal evolution in the average burst emission properties during this stage \citep{McBreen2002}. At $ t_{\rm MAD} \approx t_{\rm GRB} $, the jet power begins to decay, following a $ P(t) \sim t^{-\alpha} $ behavior with $ 1.5 \lesssim \alpha \lesssim 2 $, consistent with the observed extended emission phase \citep{Giblin2002}.

The GRB duration is set by the characteristic timescale over which the disk transitions to the MAD state ($ \phi = \phim $), i.e. $t_{\rm GRB} = t_{\rm MAD}$. From Eqs.~\eqref{eq:Mdot}, \eqref{eq:PBH} we have
\begin{equation}
  \PBH = \eta_a\dot{M}c^2 \approx \eta_a\dot{M}(1\,\s)\left(\frac{t_{\rm GRB}}{1\,\s}\right)^{-\alpha}c^2\,.
\end{equation}
Across a range of disk-producing mergers, the BH mass varies by less than a factor of $ \sim 2 $ \citep[e.g.,][]{Foucart2012a}, while the torus angular momentum is similar between events, as indicated by the nearly universal post-merger BH spin. Consequently, the radial extent of the accretion disk is expected to vary only minimally compared to its mass \citep[e.g.,][]{Fernandez2020}. This implies that $ \dot{M} \propto M_d $, leading to the GRB duration scaling with jet power as
\begin{align}\label{eq:tGRB}
    t_{\rm GRB} &\approx 1\left[\frac{\eta_a\dot{M}(1\,\s)c^2}{\PBH}\right]^{1/\alpha}\,\s\notag\\
    &\approx1\left(\frac{M_d}{10^{-2}\,\msun}\frac{5\times 10^{50}\,\erg\,\s^{-1}}{\PBH}\right)^{1/\alpha}\,\s,
\end{align}
where the normalization comes from fitting the numerical results of \citet[their Fig.~1]{Gottlieb2023e}.

\subsection{Origin of lbGRBs}\label{sec:lbgrb}

Equation~\eqref{eq:tGRB} directly links the jet power, disk mass, and GRB duration: More massive disks, which all else being equal, originate from more unequal-mass binaries\footnote{Massive disks are also predicted to form around BHs preceded by a $ \gtrsim 10\,\ms $ long-lived HMNS phase \citep[e.g.,][]{Radice2018b}.}, power longer GRBs. Numerical simulations by \citet{Gottlieb2023e} demonstrate that binary mergers which produce massive disks ($M_d \approx 0.1\,\msun$) \emph{must} generate lbGRBs ($t_{\rm GRB} \sim 10\,\s$). Otherwise, Eq.~\eqref{eq:tGRB} implies a very high required jet power, greatly exceeding observed GRB luminosities.

Other scenarios for supernova-less long GRBs have recently been revived, including WD--NS \citep{Yang2022,Sun2023} or WD--BH mergers \citep{Lee2007,Lloyd-Ronning2024}, or the AIC of a WD to form a magnetar \citep{Yi1998,Metzger2008b,Cheong2024}. However, as we discuss in \S\ref{sec:alternatives}, these alternative scenarios are challenged to explain the lbGRB properties, the presence of bright KN emission, and the total energy of the explosion as constrained by late-time radio observations. 

\subsection{Origin of sbGRBs: BH-powered vs. NS-powered jets}\label{sec:sbgrb}

\subsubsection{BH-powered jets}

Within a BH-powered jet framework, the GRB duration is determined by both the disk mass and jet power [Eq.~\eqref{eq:tGRB}]. Accepting that massive disks produce lbGRBs, one can predict the properties of shorter bursts from less massive disks knowing how the magnetic flux of the disk scales with its mass.  We consider two possible scenarios for this scaling, which roughly bracket the range of possibilities.

(i) If the \emph{dimensional} magnetic flux $ \Phi $ is the same for all disks (i.e., independent of the disk mass), then all mergers will display the same peak jet power, $ \PBH \propto \Phi^{2} \sim {\rm const}$. Eq.~\eqref{eq:tGRB} then predicts that less massive disks will produce shorter bursts, e.g. $t_{\rm GRB} \propto M_{\rm d}^{1/2}$ for $\alpha = 2$, but with similar power to lbGRBs. For example, reducing the disk mass from $ M_d \sim 10^{-1}\,\msun $ to $ M_d \sim 10^{-3}\,\msun $ would result in a GRB ten times shorter but with the same luminosity as an lbGRB.

(ii) If instead, the \emph{dimensionless} magnetic flux $ \phi $ is independent of the disk mass, then $ \PBH \sim \Phi^2 \sim M_d $ [Eq.~\eqref{eq:phi}] and hence Eq.~\eqref{eq:tGRB} predicts $ t_{\rm GRB} \sim {\rm const} $. In this scenario, a reduction of the disk mass from e.g. $ M_d \sim 10^{-1}\,\msun $ to $ M_d \sim 10^{-3}\,\msun $ would produce a GRB that is $ \sim 100 $ times fainter but of the same duration. Such a scenario of constant $ \phi $ independent of $ M_d $ is expected if the BH magnetic field $B$ is originally generated in the disk (e.g., via dynamo) in equipartition with the disk's pressure $P$, i.e. $ \PBH \propto B^2 \propto P \propto \rho \propto M_d $, where $\rho$ is the disk mass density \citep[see also][]{Lu2023}. Indeed, numerical relativity simulations demonstrate that turbulent amplification is the dominant mechanism for magnetic field enhancement in the disk, showing remarkable consistency in $ \phi $ across disk masses spanning two orders of magnitude \citep{Izquierdo2025}.

sbGRBs are observed to exhibit comparable luminosities to lbGRBs but with durations shorter by $\sim 1-3$ orders of magnitude. The first scenario, where the jet power is universal and $ t_{\rm GRB} \sim \sqrt{M_d} $, is therefore consistent with the observed differences between sbGRBs and lbGRBs. In contrast, the second scenario, where the disk mass determines the jet power rather than duration, is inconsistent with observations. However, this raises concerns, as a scenario of constant dimensionless flux is more physically motivated, given that the magnetic fields needed to power a GRB are likely generated in a post-merger disk dynamo\footnote{Alternatively, the magnetic field can be generated internally via dynamo processes in the HMNS over $ \sim 100\,\ms $ (see \S\ref{sec:NS}), a duration longer than the $ \sim 10\,\ms $, required for an HMNS to form a massive disk. Consequently, any additional contribution to the BH's magnetic field from the HMNS would necessarily be associated with massive disks, further reinforcing the dependence of $ \Phi $ on disk mass.} (the initial magnetic fields of the merging NSs are likely far too weak; {e.g., \citealt{Lorimer08}). 
In this scenario, BHs would be disfavored as the central engines of sbGRBs, leaving an NS engine as the remaining possibility.

\subsubsection{NS-powered jets}\label{sec:NS}

In NS--NS mergers with a total mass $ \Mtot \lesssim \Mthresh $, below the threshold for prompt collapse (e.g., \citealt{Shibata2005,Bauswein2013}), the remnant is an HMNS \citep[e.g.,][]{Margalit2019}, which can produce a relativistic jet. Long-lived ($ 0.1\,\s \lesssim \thmns \lesssim 1\,\s $) HMNSs, as likely formed in GW170817 \citep[][]{Metzger2018,Radice2018a}, have sufficient time to amplify magnetic fields through dynamo processes \citep[e.g.,][]{AguileraMiret2023,AguileraMiret2024,Kiuchi2024,Musolino2024}, enabling them to power sbGRBs. The burst duration is then set by the HMNS lifetime before it collapses into a BH.

The power of the NS jet, $P_{\rm NS}$, depends on its magnetic field topology, which, for an HMNS surrounded by an accretion disk, is likely to resemble a split monopole configuration \citep[e.g.,][]{Metzger2018b,Metzger2018}, for which
\begin{align}\label{eq:PNS}
    \PNS = &\frac{\pi^2 B^2\RNS^4\Omega^2 \theta_p^4}{4c}\approx 7.4\times 10^{50}\left(\frac{B}{3\times 10^{15}\,{\rm G}}\right)^2\times\notag\\& \left(\frac{\RNS}{15\,\km}\right)^4\left(\frac{\Omega}{4\times 10^3\,\s^{-1}}\right)^2\left(\frac{\theta_p}{1/3}\right)^4\frac{\erg}{\s}\,,
\end{align}
where $ \Omega \approx 4 \times 10^3 \,\s^{-1} $ and $ \RNS \approx 15\,\km $ are the characteristic angular rotation frequency and radius of the HMNS, respectively \citep[e.g.,][]{Guilet2017,AguileraMiret2023,AguileraMiret2024,Combi2023}. Here, $ \theta_p $ is the latitude on the NS surface as measured from the rotational axis above which magnetic field lines thread the relativistic jet (an estimate for the angular extent of the jet-feeding polar cap region of the NS surface is given in Appendix~\ref{sec:loading}). Given the uncertainties in Eq.~\eqref{eq:PNS}, we compare it with the jet power derived from numerical relativity simulations of magnetized post-merger NS remnants \citep[e.g.,][]{Combi2023, Kiuchi2024} and find broad consistency between them.

The greatest challenge to an HMNS-powered jet scenario is the contamination of the jet by baryons ablated from the HMNS surface due to neutrino heating by the hot remnant \citep{Thompson2001,Dessart2009,Metzger2018,Bamber2024}, which limits the jets' magnetization and, ultimately, the bulk Lorentz factor to be mildly relativistic \citep[e.g.,][]{Lee2007,Rezzolla2011,Ciolfi2019, Ruiz2019,Ruiz2020b,Ciolfi2020,Mosta2020,Sun2022,Bamber2024,deHaas2024,Musolino2024}.

In Appendix~\ref{sec:loading}, we provide an analytic estimate for the baryon feeding rate into the jet, $ \dot{M}_j $. The asymptotic Lorentz factor achievable by the jet can then be estimated as
\begin{align}\label{eq:Gamma} \nonumber
    \Gamma_\infty =&\frac{\PNS}{\dot{M}_{\rm j}c^2}\approx 70\left(\frac{B}{3\times 10^{15}\,{\rm G}}\right)^2\left(\frac{\RNS}{15\,\km}\right)^{-1/6}\times\\&\left(\frac{L_\nu}{10^{52}\,\erg\,\s^{-1}}\right)^{-17/12}\left(\frac{\epsilon_\nu}{10\,{\rm MeV}}\right)^{-10/3}\left(\frac{M}{2.7\,\msun}\right)^{2}\,,
\end{align}
where the $ \nu_{e} $ and $ \bar{\nu}_{e} $ luminosity $ L_{\nu}$ and characteristic neutrino energy $\epsilon_\nu$ are scaled to a characteristic values within the first seconds after the merger (e.g., \citealt{Dessart2009}). Equations~\eqref{eq:PNS} and \eqref{eq:Gamma} demonstrate that, for typical magnetar field strengths, HMNSs generate ultra-relativistic jets \citep[see also][]{Metzger2008b}, consistent with the power and Lorentz factor $ \Gamma \gtrsim 10 $ inferred from sbGRB observations \citep[][]{Nakar2007}. Indeed, recent high-resolution simulations by \citet{Kiuchi2024} have confirmed the presence of $ \Gamma \gtrsim 10 $ jets from magnetars.

Nevertheless, analyses of early afterglow observations of long GRBs suggest that the jet's Lorentz factor scales weakly with the GRB luminosity as $ \Gamma \sim L^{1/3} $ \citep{Hascoet2014,Ghirlanda2018}. Since sbGRBs have luminosities comparable to those of long GRBs, applying these results to sbGRB jets implies they may require $ \Gamma \gtrsim 100 $, challenging models of NS-powered jets. However, both collapsar jets and lbGRBs are likely intrinsically distinct from sbGRBs, leaving it uncertain whether these results can be applied to sbGRBs. Future observational studies of the sbGRB jet's Lorentz factor are therefore crucial for distinguishing observationally between BH- and NS-powered jets.

The NS-powered jet will shut off once the HMNS collapses into a BH. However, the newly formed BH is likely to inherit the magnetic fields of the HMNS, allowing it to launch a BZ jet instantaneously \citep{Gottlieb2024}. Assuming that magnetic poloidal flux $\Phi \simeq \pi B R_{\rm NS}^{2}$ is indeed conserved during BH formation, the change in jet power during the transition from NS-driven to BH-driven jet can be expressed as the ratio of Eq.~\eqref{eq:PNS} to Eq.~\eqref{eq:PBH}:
\begin{align}\label{eq:ratio}
    \frac{\PNS}{\PBH} \approx 15&\left(\frac{\RNS}{15\,\km}\right)^{4}\left(\frac{\Omega}{4\times 10^3\,{\s^{-1}}}\right)^{2}\left(\frac{M}{2.7\,\msun}\right)^{-2}\times\notag\\
    &\left(\frac{\theta_p}{1/3}\right)^4\left(\frac{\phim}{50}\right)^2\left(\frac{\eta_a}{0.5}\right)^{-1}\,,
\end{align}
where $ \eta_a = 0.5 $ is obtained for a spin parameter $ a = 0.72 $ typical of those found in numerical relativity simulations\footnote{Note that the final spin of the BH is not directly related to the pre-collapse angular frequency of the HMNS, $ \Omega $, via a naive application of angular momentum conservation during the collapse, $ a = I \Omega c(GM^2)^{-1} $, where $ I \approx 0.4\,M\RNS^2 $ is the moment of inertia. This is because the forming BH encompasses not only the NS but also its quasi-Keplerian envelope, which extends out to a few NS radii, carrying additional angular momentum and increasing the moment of inertia by a factor of a few.} \citep[see e.g.,][]{Kiuchi2009,Kastaun2015,Sekiguchi2016,Dietrich2017}.

The order of magnitude ratio in Eq.~\eqref{eq:ratio} suggests that a BZ jet will launch at the end of the HMNS's lifetime, but with sufficiently reduced power to signify the end of the prompt emission ($ T_{50} $; see \S\ref{sec:brightness}). The BZ jet may still contribute to the emission tail and hence can account for the extended emission observed in sbGRBs. However, we emphasize that the drop in power indicated by Eq.~\eqref{eq:ratio} is sensitive to the NS radius and angular momentum as well as the magnetic configuration on the HMNS surface, which may differ from our estimates and is likely to vary between systems. In some cases, the HMNS-powered jet may constitute an early bright phase of a longer GRB powered by a BH-driven jet.

\section{GRB-Kilonova connection}\label{sec:kn}

We extend the theoretical framework linking GRB observations and binary populations to include KNe, aiming to establish observational connections between GRB classes and KN signal properties. We first outline the general principles and then apply them to GRB--KN observations.

\subsection{Kilonova luminosity}\label{sec:brightness}

Although the KN luminosity depends on several factors, the dominant one is the ejecta mass, composed of winds from the accretion disk and dynamical ejecta from the merger process. Simulations tracking the evolution of post-merger accretion disks endowed with strong (weak) magnetic fields find that a substantial fraction of the disk mass is ejected, $ M_{\rm ej} \approx 0.4~(0.3)\,M_d $ \citep[e.g.,][]{Christie2019,Fernandez2019,Nedora2021a,Gottlieb2023b,Kiuchi2023b}.  Since $ \phi $ needs to be initially low to match observed lbGRB luminosities in massive disks -- and is likely consistent across different disk masses -- we adopt an ejecta mass of $ M_{\rm ej} = 0.3\,M_d $. In a suite of NS--NS simulations by \citet{Radice2018b,Radice2018a}, the disk winds consistently outweigh the dynamical ejecta, with $ 0.3\,M_d \gg M_{\rm dyn} $, except in a few cases involving a massive binary with soft equations of state (EoS), resulting in low disk masses, $ M_d < 10^{-3}\,\msun $ and $ M_{\rm ej} \sim 10^{-4}-10^{-3}\,\msun $. But in this regime, the KN signal is likely too faint for detection anyway, suggesting that when a KN signal is detectable, the disk winds dominate the ejecta mass (e.g., \citealt{Wu+16,Siegel22}). In summary, the KN luminosity scales most sensitively with the ejecta mass, which in turn scales with the disk wind mass, and ultimately with the disk mass.

lbGRBs arise from BHs with massive disks of $ M_d \sim 0.1\,\msun $, suggesting that all lbGRBs must be accompanied by a bright KN. Massive disks can form in moderate mass-ratio BH--NS mergers with high pre-merger BH spin \citep{Foucart:2012vn,Kyutoku:2015gda}, in NS--NS mergers with unequal mass ratios, or in the presence of a $ \gtrsim 10\,\ms $ long-lived HMNS. The different scenarios would significantly affect the ejecta composition, highlighting the importance of considering the KN color as well (\S\ref{sec:color}).

For sbGRBs, we consider both BH-driven and NS-driven jets (\S\ref{sec:sbgrb}). Starting with the BH case, Eq.~\eqref{eq:tGRB} can be rearranged to connect the ejecta mass to the jet duration and energy $ E_j = \PBH t_{\rm GRB} $,
\begin{equation}
    M_d \approx 10^{-2}\left(\frac{E_j}{5\times 10^{50}\,\erg}\right)\left(\frac{t_{\rm GRB}}{1\,\s}\right)^{\alpha-1}\,\msun\,.
\end{equation}
The jet energy can also be related to the observed isotropic equivalent $ \gamma $-ray energy, $ \Eisog $, through
\begin{equation}
    \Eisog = \frac{\epsilon_\gamma}{f_b}E_j = 10f_{-1} E_j\,,
\end{equation}
where $ f_{-1} \equiv 0.1\epsilon_\gamma/f_b $, $ 0.01 \leq f_b \leq 0.11 $ is the beaming fraction, and $ 0.15 \leq \epsilon_\gamma \leq 0.5 $ represents the radiative efficiency of the $ \gamma $-ray emission \citep[see discussion in][]{Gottlieb2023e}. 
Assuming $ M_{\rm ej} \approx 0.3\,M_d $ and taking $ t_{\rm GRB} \approx T_{50} $ given that the prompt and extended emissions carry comparable amounts of energy \citep{Kaneko2015,Zhu2022}, the ejecta mass is estimated as
\begin{equation}\label{eq:Mej}
    \mej \approx 10^{-3}f_{-1}^{-1}\left(\frac{\Eisog}{2\times 10^{51}\,\erg}\right)\left(\frac{T_{50}}{1\,\s}\right)^{\alpha-1}\,\msun\,.
\end{equation}
Given that lbGRBs and sbGRBs exhibit comparable $\gamma$-ray luminosities and hence similar jet power $ \PBH $, the ejecta mass in this scenario scales as $ M_{\rm ej} \sim M_d \sim t_{\rm GRB}^\alpha \sim t_{\rm GRB}^{2}$ [Eq.~\eqref{eq:tGRB}]. Since the $ T_{50} $ of lbGRBs is $ \sim 1.5 $ orders of magnitude longer than that of sbGRBs, then, if both lbGRBs and sbGRBs are powered by BHs, Eq.~\eqref{eq:Mej} indicates that the KNe accompanying lbGRBs would be $ \sim 1000$ times brighter than those associated with sbGRBs. Consequently, sbGRB KNe would need to be 1000 times closer for detection. However, given that the rates of lbGRBs and sbGRBs are comparable (Levan et al. 2025), this implies that, in practice, \emph{if all sbGRBs were BH-powered, their KNe would be too faint to detect}.

On the other hand, if NSs serve as the central engines of sbGRBs, the jet duration is instead determined by the HMNS lifetime. Longer-lived HMNSs might produce higher ejecta mass, but the increase is moderate, particularly along the polar axis \citep{Lippuner2017,Fujibayashi2018,Darbha2020,Fujibayashi2023}. Therefore, if sbGRBs are powered by NSs, no strong correlation between the GRB duration and the KN luminosity is expected.

\subsection{Kilonova color}\label{sec:color}

\begin{figure*}
\includegraphics[trim={0cm 1.5cm 0 1.5cm},scale=0.64]{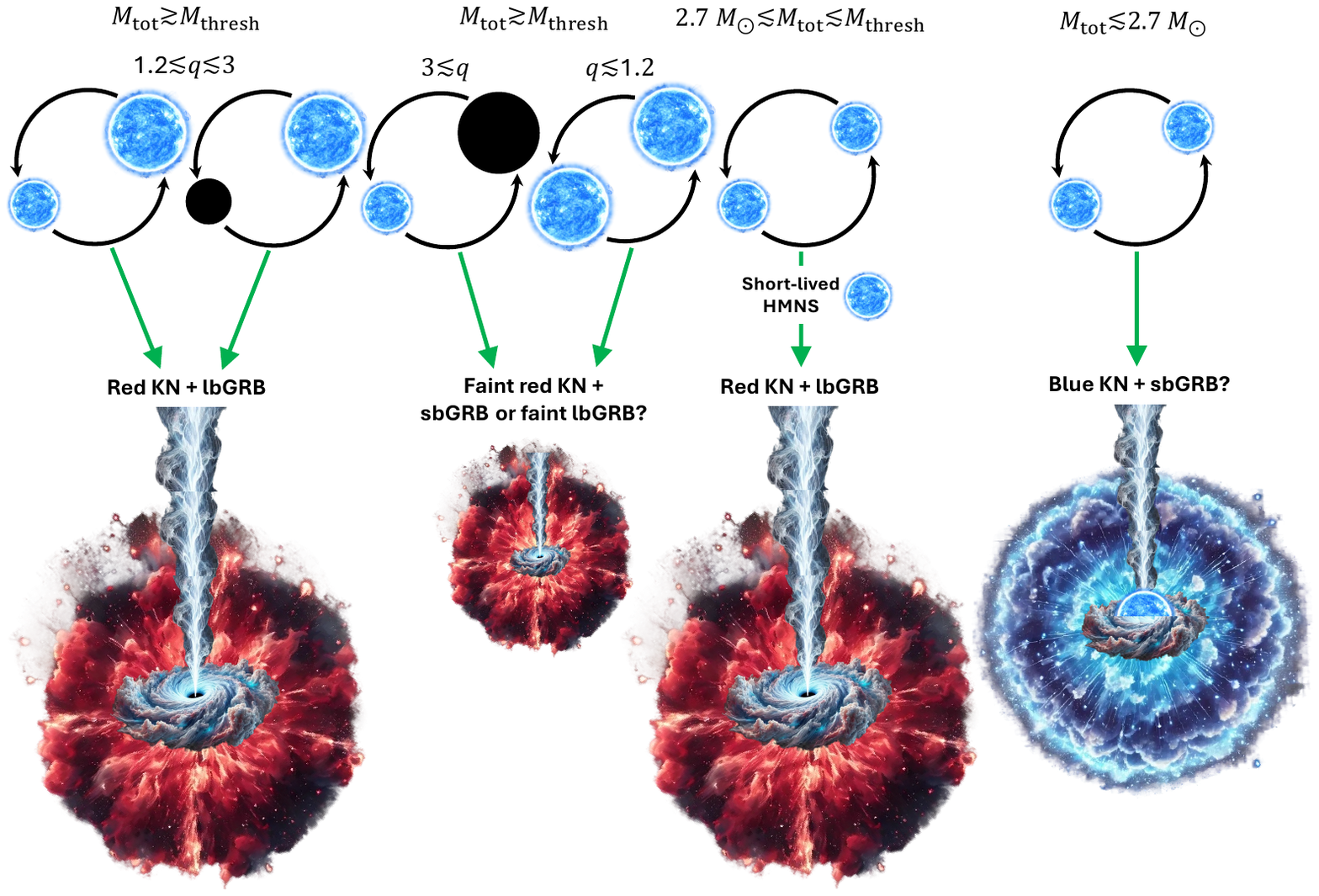}
\caption{How the outcomes of compact binary mergers connect to their GRB and KN properties.  From left to right: massive binary systems with $ \Mtot \gtrsim \Mthresh $ promptly collapse into BHs, powering a relatively red KN. If the binary mass ratio is moderate ($ 1.2 \lesssim q \lesssim 3 $), a substantial accretion disk forms around the BH, producing an lbGRB and bright KN emission. In contrast, mergers with near-equal mass NSs or a BH--NS system with a high mass ratio will result in a smaller disk and a fainter KN signal.
If $ \Phi $ remains consistent across mergers with varying disk masses, the GRB power will be similar to lbGRBs, implying an sbGRB for less massive disks. If $ \phi $ remains constant across different disk masses, as seen in simulations \citep{Izquierdo2025}, these mergers yield a fainter lbGRB. When $ \Mtot \lesssim \Mthresh $, the remnant is an HMNS. A short-lived HMNS ($ \Mtot \gtrsim 2.7\,\msun $) will collapse into a BH with a massive disk, resulting in GRB and KN signals similar to the leftmost scenario. However, a less massive HMNS that survives for $ \sim 1\,\s $ (rightmost scenario) may power an sbGRB accompanied by purple KN emission.
\label{fig:sketch}
}
\end{figure*}

We consider the KN color as a function of the ejecta electron fraction, where $ \Ye \lesssim 0.2 $ corresponds to a red KN, $ 0.2 \lesssim \Ye \lesssim 0.3 $ to a purple KN, and $ \Ye \gtrsim 0.3 $ to a blue KN (e.g., \citealt{Tanaka2017}). Massive disks around BHs powering lbGRBs are expected to produce more neutron-rich ejecta from the disk with $ \Ye \lesssim 0.2 $, especially in the presence of strong magnetic fields associated with higher viscosity \citep{Fernandez2019,Kawaguchi2024}, giving rise to a bright red optical/infrared KN signal \citep[e.g.,][]{Siegel2017,Kasen2017}. Such massive disks can form from unequal-mass binary mergers, producing substantial isotropic shock-heated ejecta that contribute a bluer component to the emission in these events \citep{Oechslin2007,Sekiguchi2016,Radice2018b}. Alternatively, massive disks may also form following the collapse of a $ \gtrsim 10\,\ms $ long-lived HMNS. In such cases, neutrino irradiation from the HMNS may dominate the blue component, with the $ \Ye $ increasing with $ \thmns $.

If sbGRBs are powered by BHs with less massive disks, the KN may receive a significant contribution from the dynamical ejecta, as discussed in \S\ref{sec:brightness}. The composition of the dynamical ejecta is split between red tidal tail ejecta and blue shock-heated ejecta, which depends strongly on the EoS \citep[e.g.,][]{Sekiguchi2015,Sekiguchi2016,Radice2018b}. As a result, the KN in this case could be either bluer or redder than those associated with lbGRBs. However, as argued in \S\ref{sec:brightness}, such low-mass disks are not expected to produce a KN luminous enough for detection.

Long-lived HMNSs irradiate the disk with neutrinos, which act to increase the electron fraction of the disk outflows \citep[e.g.,][]{Metzger2009,Darbha2010,Metzger2014,Nedora2021a,Radice2024}. If sbGRBs are powered by HMNSs, their characteristic duration, $ T_{90} \approx 0.8\,\s $ \citep[e.g.,][]{Tarnopolski2016}, suggests $ \thmns \sim 1\,\s $. Such long-lived HMNSs would persist throughout most of the disk wind ejection, maintaining a high electron fraction of $ \Ye \gtrsim 0.3 $ due to strong neutrino irradiation, especially along the poles \citep[e.g.,][]{Metzger2014,Perego2014,Lippuner2017,Fujibayashi2018}, most relevant to KNe associated with GRB-producing face-on mergers\footnote{While off-axis events, such as GRB 170817A, may produce detectable $\gamma$-ray emission, they are intrinsically much fainter and can only be detected at very close distances. As a result, they do not significantly impact a large sample population.}. However, regardless of the central engine's fate, the late-time ejecta from the disk is expected to possess a similar electron fraction, $ \Ye \sim 0.3 $, due to the weak freeze-out that occurs as the disk viscously spreads, the temperature drops, and free particles combine into alpha-particles \citep[e.g.,][]{Lee2009,Metzger2009,Fernandez2013,Shibata2021,Shibata2021b,Fahlman2022,Kawaguchi2024}.

Figure~\ref{fig:sketch} illustrates the various potential outcomes of different binary systems. lbGRBs are produced by massive accretion disks around BHs and are accompanied by bright red KNe. These disks can form either from binary systems with mass ratios of $ 1.2 \lesssim q \lesssim 3 $ (leftmost scenarios) or from short-lived HMNS with $ 2.7\,\msun \lesssim \Mtot \lesssim M_{\rm thresh} $, which enriches the disk mass prior to collapse. In contrast, binaries with near-equal mass or extreme mass ratios that do not go through a metastable NS phase result in a less massive post-merger disk. This may produce a similarly red KN, but it is likely too faint to be detected. The resulting jet could be a standard sbGRB or a dimmer lbGRB, depending on how the magnetic flux scales with disk mass. If sbGRBs are powered by long-lived HMNSs (rightmost scenario), the corresponding KN should differ from those associated with lbGRBs, exhibiting purple KN emission. These predictions offer a valuable test for using GRB-associated KN observations to validate our massive BH disk model for lbGRBs and infer the central engine of sbGRBs.

\subsection{Observations and implications for central engines}\label{sec:observations}

\citet{Rastinejad2024} recently conducted multi-component modeling of eight KNe (following the same division by $ Y_e $ as presented here) associated with four sbGRBs, three lbGRBs, and GW170817 \citep[where the $ \gamma $-rays did not originate from the jet; see, e.g.,][]{Gottlieb2018b, Mooley2018b, Matsumoto2019}. Table~\ref{tab:modeling} summarizes their results for the KN ejecta and its connections to the prompt GRB emission. 

\begin{table}[h]
\setlength{\tabcolsep}{2.8pt}
    \centering
    \renewcommand{\arraystretch}{1.2}
\begin{tabular}{|c|c c c | c c c c c|}
\hline 
    GRB & $ T_{90} $ & $ T_{50} $ & $ \Eisog $ & $ M_b $ & $ M_p $ & $ M_r $ & $ M_{\rm ej} $ & $ E_{\rm ej} $ \\ 
     & [s] & [s] & $ [10^{50}\,\erg] $ & \multicolumn{4}{c}{[$ 10^{-2}\,\msun $]} & $ [10^{50}\,\erg] $ \\\hline 
    170817A & 2.00 & 1.0 & 0.06 & 0.4 & 1.9 & 5.2 & 7.5 & 23 \\\hline
    050709 & 0.07 & 0.03 & 3.1 & 0.3 & 2.9 & 1.5 & 4.7 & 5.6\\
    130603B & 0.18 & 0.06 & 29 & 0.6 & 7.5 & 2.3 & 10.4 & 15\\
    160821B & 0.48 & 0.28 & 0.5 & 0.3 & 1.1 & 1.1 & 2.5 & 4.2\\
    200522A & 0.62 & 0.38 & 8.1 & 4.6 & 2.0 & 1.9 & 8.5 & 32\\\hline
    060614 & 109 & 43.2 & 51 & 1.4 & 2.9 & 14.6 & 18.9 & 68\\
    211211A & 50.7 & 21.2 & 172 & 0.7 & 1.0 & 13.0 & 14.7 & 97\\
    230307A & 35.0 & 20.0 & 290 & 1.2 & 2.5 & 5.4 & 9.1 & 26\\\hline
\end{tabular}

    \caption{Summary of observed GRB and KN modeling from \citet{Rastinejad2024}: $ T_{90}~(T_{50}) $ is the time containing 90\% (50\%) of the prompt emission, $ \Eisog $ is the isotropic equivalent $ \gamma $-ray energy, $ M_b, M_p, M_r $ represent the blue, purple, and red ejecta masses, respectively, $ M_{\rm ej} $ is the total ejecta mass, and $ E_{\rm ej} $ is the total kinetic energy of the ejecta. For GRBs 170817A, 050709 and 230307A, we infer $ T_{50} $ by extrapolating the $ T_{90}/T_{50} $ ratio from other GRBs.}
        \label{tab:modeling}
\end{table}

Whereas \citet{Rastinejad2024} observed a moderate range of total ejecta masses\footnote{This is consistent with estimates derived solely from turbulent mixing constraints of r-process elements \citep{Kolborg2023}.}, the red and blue ejecta masses varied significantly, suggesting a diverse range of binary populations. Another apparent trend is that KNe associated with lbGRBs show significantly more massive red ejecta than those associated with sbGRBs (see also Figure~\ref{fig:T50_Mej}). Based on the above discussion, this supports our conclusion that massive BH disks power lbGRBs. Such massive disks, which either arise from an unequal mass-binary or a $ \gtrsim 10\,\ms $ long-lived HMNS, also typically generate larger amounts of high$-Y_{e}$ blue ejecta, consistent with the trends in Tab.~\ref{tab:modeling}.

\begin{figure}
\includegraphics[scale=0.18]{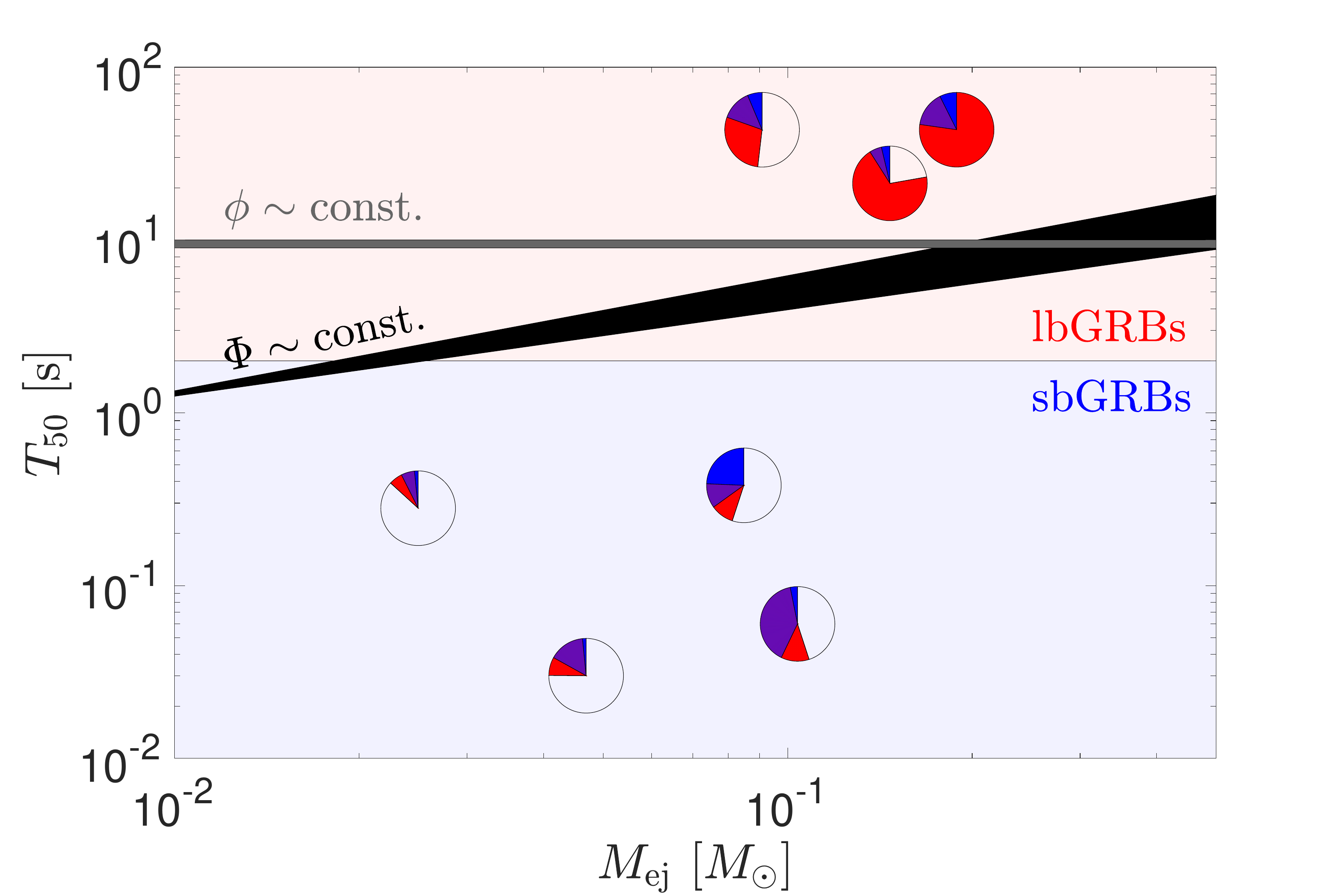}
\caption{Expectations for the relationship between KN ejecta mass and GRB duration in BH-powered jet models for lbGRBs (shaded red area) and sbGRBs (shaded blue area). The pie charts represent the red, purple, and blue ejecta fractions of $ M_{\rm ej} $, normalized to the maximum ejecta mass of the KN associated with GRB 060616. This highlights that lbGRBs are consistently associated with a more substantial red KN component, while sbGRBs show relatively bluer and more purple emission. The black region shows the trend predicted in a scenario where the (dimensional) magnetic flux $\Phi$ on the BH is the same for all mergers. In contrast, the gray line shows the trend predicted if the dimensionless flux $\phi$ is instead held fixed. The normalization of the trends is arbitrary, while their thickness follows from varying the power-law index of the BH accretion rate $\dot{M} \propto t^{-\alpha}$ in the range $ 1.5 \lesssim \alpha \lesssim 2 $. Neither model aligns with the much steeper observed relationship between the $ T_{50} $ of the GRB and the $ M_{\rm ej} $ of the KN in the GRB--KN sample modeled by \citet{Rastinejad2024} (Tab.~\ref{tab:modeling}). Both this discrepancy and the distinct KN color distributions suggest that sbGRBs are powered by HMNSs.
\label{fig:T50_Mej}
}
\end{figure}

Comparing the properties of the KNe associated with sbGRBs versus lbGRBs provides additional constraints on the central engine behind sbGRBs. Considering a scenario in which both lbGRBs and sbGRBs are powered by BH jets (Sec.~\ref{sec:lbgrb}), Fig.~\ref{fig:T50_Mej} compares the observed KNe ($ \mej $ and color distribution) with the $ T_{50} $ of the associated GRBs. The lines delineate the expected correlations between $ T_{50} $ and $ \mej $ using Eq.~\eqref{eq:Mej} under the assumptions of either constant magnetic flux $ \Phi $ (black region) or dimensionless magnetic flux $ \phi $ (gray line). If $ \phi $ were independent of disk mass, all GRBs should exhibit similar durations, precluding the very existence of sbGRBs. If $ \Phi $ remains consistent across disk masses, then sbGRBs and lbGRBs should exhibit comparable luminosities, but their durations would scale as $ \mej \sim t_{\rm GRB}^\alpha \sim t_{\rm GRB}^{2}$. Given the wide range of GRB durations, this would imply orders of magnitude variations in the KN luminosity observed. Had this correlation existed, we would expect an observational bias toward detecting KNe associated with lbGRBs, i.e., the KNe accompanying sbGRBs should not be detectable. However, this conflicts with the inferred KN ejecta mass differing by less than an order of magnitude between sbGRBs and lbGRBs (Tab.~\ref{tab:modeling}). Neither scenario in which sbGRBs are powered by BHs therefore appears consistent with observations. Moreover, \citet{Gottlieb2023e} demonstrated that BHs with disks of $ M_d \gtrsim 0.1\,\msun $, as inferred for all mergers in the sample, will inevitably power lbGRBs, thereby ruling out a BH-engine model for sbGRBs.

Fig.~\ref{fig:T50_Mej} and Table~\ref{tab:modeling} further show that KNe associated with sbGRBs are comparatively bluer than those associated with lbGRBs, suggesting that neutrino irradiation from an NS remnant may increase the ejecta $ \Ye $ in sbGRBs. This also supports HMNSs, rather than BHs, being the central engines of sbGRBs. For HMNS-powered jets, the dependence on disk mass is weak, supporting our model and the expected association of longer-lived HMNSs with more equal-mass binaries. This suggests that, as expected, $ \phi $ is largely independent of disk mass, with BH-hosting less massive disks responsible for powering faint lbGRBs rather than standard sbGRBs; such systems are therefore more challenging to detect in both $\gamma$-rays and optical wavelengths given their low-mass KNe.

\section{Conclusions}\label{sec:summary}

\begin{figure*}[]
\centering
\includegraphics[scale=0.7]{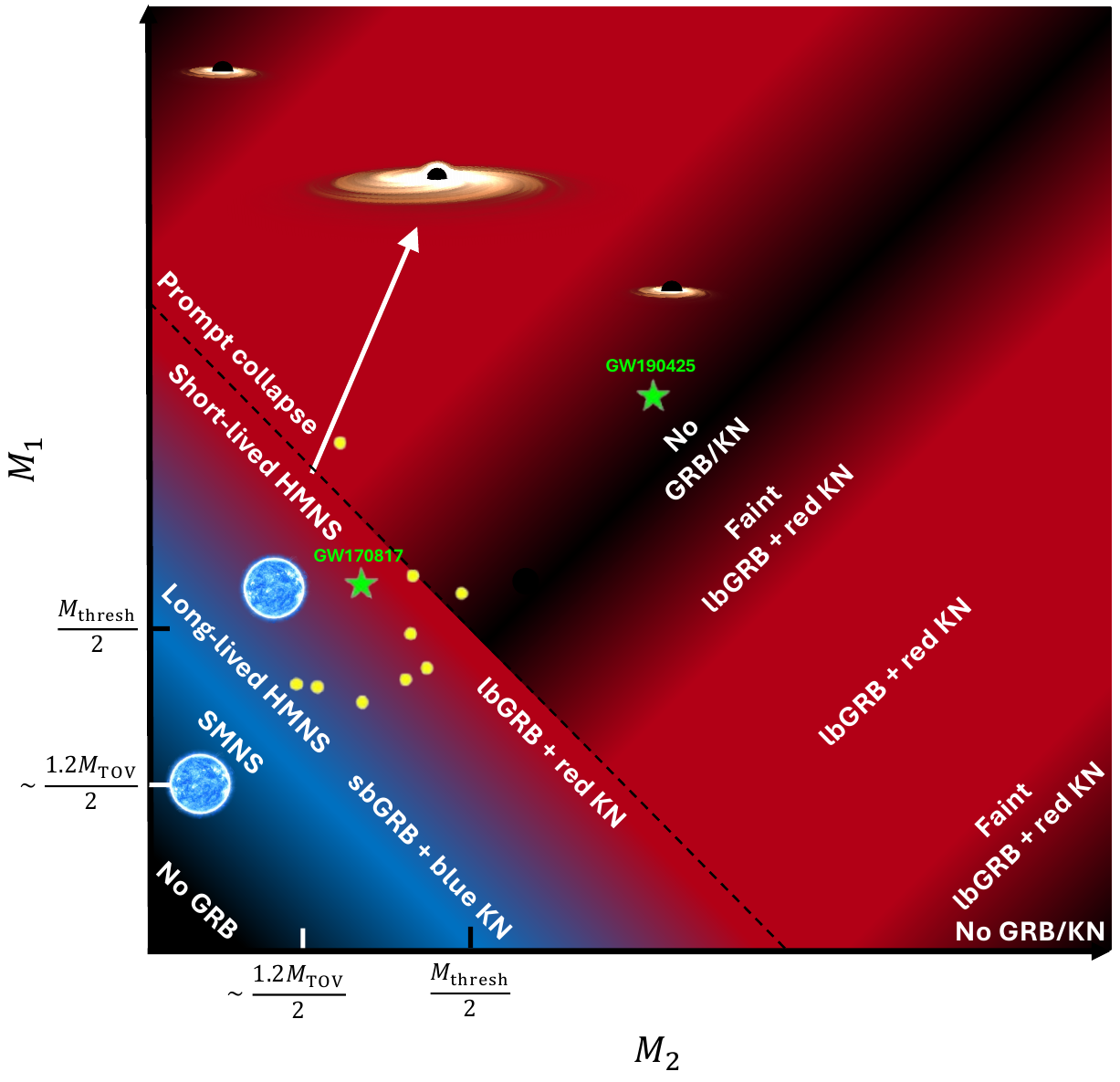}
\caption{The GRB and KN outcomes of compact object mergers with masses $ M_1 $ and $ M_2 $. The top-left half depicts the resultant central engine, while the bottom-right half shows the GRB and KN properties. Yellow dots indicate the Galactic NS--NS population that will merge within Hubble time \citep{Chu2022}, green stars indicate NS--NS mergers from \citet{Abbott2019b,Abbott2020b}, assuming $ M_{\rm TOV} = 2.05\,\msun $ and $ M_{\rm thresh} = 2.8\,\msun $. If $ \Mtot \sim 1.2\,M_{\rm TOV} $, the merger produces a long-lived ($ 100\,{\rm ms} \lesssim \thmns \lesssim 1\,{\rm s} $) remnant, potentially through a supramassive NS (SMNS) phase \citep{Margalit2022}, powering an sbGRB with a blue KN (dominated by a blue-purple component). More massive HMNSs collapse rapidly ($ 10\,{\rm ms} \lesssim \thmns \lesssim 100\,{\rm ms} $), forming a BH with a massive disk ($ M_d \gtrsim 0.1\,\msun $) capable of powering an lbGRB with a red KN (dominated by the red component) -- the area just below the dashed line. The dashed line marks the binary mass threshold above which an NS--NS merger results in a prompt collapse. Equal-mass or extreme mass-ratio binaries with $ M > \Mthresh $ do not form a disk, preventing the production of a jet or KN (black areas). For binaries with $ 1.2 \lesssim q \lesssim 3 $, a substantial disk allows for an lbGRB and a bright red KN. Between these two regimes, less massive disks produce fainter lbGRBs and KN emission.
\label{fig:BNB_pop}
}
\end{figure*}

In this paper, we expand the theoretical framework that links binary merger populations to lbGRBs and sbGRBs \citep{Gottlieb2023e} to also include their connection with KNe. lbGRBs powered by massive disks around BHs are expected to be associated with bright, red KNe. If sbGRBs were powered by BHs, they would be accompanied by significantly fainter, red KNe. However, we demonstrated that HMNSs, given characteristic magnetic field strengths and spin rates, could also power sbGRBs with the luminosity and Lorentz factors inferred from observations, though future studies are essential for more precise constraints on the latter. HMNS-powered sbGRBs would be accompanied by a bluer KN, making KN observations key for identifying the central engines of GRBs.

Recent modeling of GRB--KN observations by \citet{Rastinejad2024} shows that lbGRBs are accompanied by luminous red KN signals, supporting the prediction that lbGRBs are powered by BHs with massive disks. The study further suggests that sbGRBs are associated with KNe that are nearly as luminous, implying disk masses comparable to those accompanying lbGRBs. Since BHs with massive disks must power lbGRBs, this finding challenges the possibility of BH engines for sbGRBs. Indeed, the KNe associated with sbGRBs exhibit a distinctly bluer hue. Since all lbGRB-associated KNe (which are brighter) are observed at closer distances than those associated with sbGRBs \citep{Rastinejad2024}, it is plausible that the current sample of sbGRBs is representative of the overall population (i.e., no observational bias exists against detecting sbGRBs with fainter KNe). This suggests that HMNSs serve as the central engines of sbGRBs, implying that sbGRBs arise from NS--NS mergers. Within this framework, all BHs power lbGRBs, consistent with turbulent dynamo field amplification that generates a uniform $ \phi $ value across varying disk masses. This suggests that BH--NS mergers do not produce sbGRBs but instead contribute to the population of lbGRBs accompanied by red KNe.

Figure~\ref{fig:BNB_pop} summarizes the outcomes of GRBs and KNe from different mergers. The upper-left half illustrates the central engine formed while the lower-right half outlines the associated GRB and KN properties. Binary mergers that give rise to long-lived HMNS generate sbGRBs accompanied by blue KNe. Shorter-lived HMNSs may collapse into BHs with substantial accretion disks, producing lbGRBs paired with red KNe. Systems above the dashed line correspond to prompt collapses and BH--NS binaries. Mergers with moderate mass ratios $ 1.2 \lesssim q \lesssim 3 $ can form massive disks, similarly resulting in lbGRBs and bright red optical/infrared KNe. The black areas correspond to equal-mass or extreme mass-ratio binaries with $ \Mtot > M_{\rm thresh} $, for which disk formation is suppressed, producing little or no jet or KN production. In between these regimes, less massive disks lead to fainter lbGRBs and weaker KN emission.

Yellow dots represent the Galactic NS--NS population that will merge within Hubble time, assuming $ M_{\rm TOV} = 2.05\,\msun $ and $ M_{\rm thresh} = 2.8\,\msun $. If redshift evolution is weak \citep{Roy2024} and hence the Galactic population is reflective of the cosmic merger population, this suggests that the majority of massive BH disks form via short-lived HMNSs rather than high mass-ratio binaries ($ q \gtrsim 1.2 $). Consequently, GRBs 211211A and 230307A are likely to have originated from systems producing short-lived HMNSs. Green stars denote the NS--NS mergers detected by LIGO--Virgo--KAGRA. GW190425 may have produced faint or no electromagnetic counterparts. Since the jet in GW170817 was not observed during the prompt phase, Fig.~\ref{fig:BNB_pop} suggests it may have appeared as an lbGRB for on-axis observers -- indeed, its KN properties most closely resemble those of the KN associated with GRB 230307A.

Using the theoretical framework presented here, \citet{Perna2024} compared the inferred sbGRB and lbGRB observed rates with the Galactic binary NS population and found evidence for an EoS harder than that corresponding to $ M_{\rm thresh} = 2.8\,\msun $. For a harder EoS, the diagonal dashed line shifts toward the top-left, implying that all Galactic NS--NS binaries merging within a Hubble time will undergo an NS phase before collapsing into a BH.

\subsection{Challenges with alternative models}\label{sec:alternatives}

Since massive disks must power lbGRBs with bright red KNe, this appears to be the most natural explanation for the origin of lbGRBs, reducing the necessity for alternative models. Any additional formation channels would only add to the population already produced by massive disks. Moreover, alternative models face major challenges, as they diverge substantially from both KN and GRB observations.

We first discuss the recently popular scenario proposing a magnetar origin for lbGRBs, following a WD--NS merger \citep{Yang2022,Sun2023} or a WD--BH merger \citep{Lee2007, Lloyd-Ronning2024}. Recent simulations of WD--NS mergers by \citet{Moran-Fraile2024} show very long activity timescales $ t \gg t_{\rm GRB} $, with jet power $ P_j \sim 10^{47}\,\erg\,\s^{-1} $, far below typical GRB luminosities (see also \citealt{Metzger2012,Fernandez2019}). Even in scenarios where long-lived magnetars could theoretically power GRBs, these have been largely disfavored due to the lack of evidence for significant rotational energy injection, as indicated by KN modeling and (lack of) late-time radio emission \citep{Metzger2014b, Horesh2016,Schroeder2020,Beniamini2021,Wang2024}. The accretion torii produced by WD--NS or WD--BH mergers can generate large quantities of wind ejecta (e.g., \citealt{Metzger2012}). However, the maximum accretion rates achieved in these disks of $ \dot{M} < 10^{-3}\,\msun\,\s^{-1} $ (e.g., \citealt{Metzger2012,Margalit2016,Moran-Fraile2024}), are far too low to neutronize the inflowing gas, which is a prerequisite for generating neutron-rich disk winds capable of powering the red KN emission \citep[see also][]{Fernandez2019b}.

AIC events \citep{Yi1998,Metzger2008b,Cheong2024} that produce a rapidly rotating NS may drive relativistic jets if significant magnetic field amplification could occur through a dynamo process, forming a magnetar. To generate neutron-rich ejecta, the magnetar must rotate near break-up velocity (e.g., \citealt{Dessart2006,Metzger2008b}), implying that the WD must spin up by several orders of magnitude during the AIC. However, similar to the WD--NS merger case, the large rotational energy of such a millisecond magnetar $\gtrsim 3\times 10^{52}\,\erg$ necessarily transferred to the environment in such an event conflicts with the GRB afterglow or KN kinetic energies (Table \ref{tab:modeling}).

\acknowledgements
We thank Kenta Kiuchi, Eliot Quataert, Sho Fujibayashi, Carlos Palenzuela Luque, and Jillian Rastinejad for helpful discussions.
O.G. is supported by the Flatiron Research Fellowship.  B.D.M. is supported by the National Science Foundation (grant numbers AST-2009255, AST-2406637), Fermi Guest Investigator Program (grant number 80NSSC22K1574), and the Simons Investigator Program (grant number 727700).  The Flatiron Institute is supported by the Simons Foundation.
E.R.-R. acknowledges support from the Heising-Simons Foundation and the National Science Foundation (2150255 and 2307710). 
F.F. is supported by the Department of Energy, Office of Science, Office of Nuclear Physics, under contract number DE-AC02-05CH11231, by NASA through grant
80NSSC22K0719, and by the National Science Foundation through grant AST-2107932.
This work was performed in part at Aspen Center for Physics, which is supported by National Science Foundation grant PHY-2210452.

\bibliography{refs}

\appendix

\section{Baryon loading estimate in the jet}\label{sec:loading}

For a slowly rotating, unmagnetized NS which emits electron neutrinos and electron antineutrinos of luminosity $L_\nu $, the total mass-loss rate due to neutrino heating is given by \citep[e.g.,][]{Qian1996}
\begin{equation}\label{eq:Mdotout}
    \dot{M}_{\rm iso} \approx 8\times 10^{-5} \left(\frac{L_\nu}{10^{52}\,\erg\,\s^{-1}}\right)^{5/3}\left(\frac{\epsilon_\nu}{10\,{\rm MeV}}\right)^{10/3}\left(\frac{M}{2.7\,\msun}\right)^{-2}\left(\frac{\RNS}{15\,\km}\right)^{5/3}\frac{\msun}{\s}\,.
\end{equation}
For a rotating NS, Eq.~\eqref{eq:Mdotout} remains valid as the isotropic mass-loss rate along open field lines provided that the (exponential) centrifugal enhancement of $\dot{M}_{\rm iso}$ can be neglected. Centrifugal enhancements are small where the surface rotational velocity is less than the surface sound speed \citep{Metzger2007}, i.e.~ for latitudes $\theta_{\rm p}$ that obey
\begin{equation}
    v_\varphi \approx \RNS\Omega \sin\theta_{\rm p} < c_s \approx \left(\frac{kT_\nu}{m_p}\right)^{1/2}\approx \left(\frac{2k^4L_\nu}{7\pi\sigma_{\rm SB} m_p^4 R_{\rm NS}^2}\right)^{1/8}\,,
\end{equation}
where $ \sigma_{\rm SB}$ is the Stefan–Boltzmann constant, and $ T_\nu \approx (2L_{\nu}/7\pi \sigma_{\rm SB} R_{\rm NS}^{2})^{1/4}$ is the neutrinosphere temperature assuming a Fermi-Dirac blackbody spectrum. This condition $v_\varphi \approx c_{\rm s}$ can be used to define the opening angle of a minimally baryon-loaded polar funnel,
\begin{equation}
    \theta_{\rm p} \approx \frac{1}{3}\left(\frac{L_\nu}{10^{52}\,\erg\,\s^{-1}}\right)^{1/8}\left(\frac{\RNS}{15\,\km}\right)^{-5/4}\left(\frac{\Omega}{4\times 10^3\,\s^{-1}}\right)^{-1}\,{\rm rad}\,,
\end{equation}
which naturally defines the extent of the polar magnetic flux feeding the HMNS jet. The baryon feeding rate into the jet is thus given by $\dot{M}_{\rm j} \approx \dot{M}_{\rm iso}f_{\rm open}$, where $f_{\rm open} \approx \theta_{\rm p}^{2}/2$ is the fraction of the HMNS surface threaded by the magnetic field lines of the jet. 

\end{document}